\documentclass[12pt,a4paper]{article}
\usepackage{amsfonts}
\usepackage{graphicx}
\usepackage{amsmath}
\usepackage{amssymb}

\begin{document}

\title{On interaction in extended particle model on $\left(  M^{4}\times
M^{4}\right)  \otimes Z_{4}$}
\author{A. Smida\thanks{email: asmida@wissal.dz}, M. Hachemane\thanks{email:
mhachemane@wissal.dz}, A.H. Hamici et R. Djelid.\\Facult\'{e} de Physique,\\USTHB, B.P. 32 El-Alia Bab-Ezzouar 16111,\\Algiers Algeria.}
\maketitle

\begin{abstract}
A $\left(  M^{4}\times M^{4}\right)  \otimes Z_{4}$ model, describing an
extended particle composed of two local modes and represented by a field
$\psi\left(  x,\xi;z\right)  $, is formulated in its most general form
($\left(  x,\xi;z\right)  \in\left(  M^{4}\times M^{4}\right)  \otimes Z_{4}%
$). The $z$ argument specifies whether the particle is observable,
unobservable, or partially observable (the latter case appears in two forms).
In this four-sheeted structure, each sheet posses its own symmetry localized
with respect to both space-times $M^{4}$ inducing thereby connections in the
continuous directions. Connections in the discrete direction describe
transitions between observable, unobservable, and partially observable states.
Curvatures and propagators are determined.

\end{abstract}

\section{Introduction}

The present work is a reconsideration of a previous one,\cite{Smida 2000}
aiming at a description of conversion of external modes into internal ones,
and vice-versa, in a geometro-differential conception of extended particles.

Initially,\cite{Smida 1995} this conception was based on a fiber bundle
$E^{D}(M,H^{D},U^{D}(G^{\prime}))$. The base space $M$ is ordinary space-time
which may have a $G$ symmetry or be curved. The typical fiber $H^{D}$ is a
Hilbert space carrying an induced representation\cite{Mensky 1976}
$U^{D}\left(  G^{\prime}\right)  $ of the internal symmetry group $G^{\prime}%
$. The particle extension stems from the fact that the latter is no more
represented by a point $x\in M$, but by a function $\Psi\in H^{D}$ which
depends on another spatiotemporal variable $\xi$ with the $G^{\prime} $
symmetry. The function $\Psi$ is not the probabilistic wave function but
describes the intrinsic properties of the particle (of which $x$ is a partial
representation).\cite{Destouches 1956} The role of probability amplitude,
played by the quantum mechanics wave function $\psi\left(  x\right)  $, is
guaranteed by a functional\cite{Destouches 1956} $X[\Psi,t]$ in this
conception. The function $\Psi_{x}\left(  \xi\right)  $ must be treated
according to some realistic model, we assumed that it describes a quantum mode
localized at $\xi$ for a particle localized at $x$ (from a partial
standpoint). The particle is composed then of two modes.\cite{Smida 1995}

The probabilistic functional $X[\Psi](x,\xi)=\psi(x,\xi)$ has been chosen as a
bilocal function representing an external quantum mode localized at $x$ and an
internal one localized at $\xi$. This function was quantized by applying the
induced representation method to both the external and internal symmetries.
When interaction is absent, the induced representation method leads to a
propagator which is a product of the propagators of each local mode. The
external mode propagation is determined by a transition from external
configuration induced representation to the external momentum one and back to
the external configuration representation. Internal mode propagation is
realized in the internal spaces. If an interaction is represented by a gauge
field (a connection in the fiber), the semigroup induced representations lead
to a path integral propagator.\cite{Mensky 1983, Smida 1998}

The assumption that the external mode may transit via internal momentum space
and the generalization of this idea to the possibility of transitions between
external and internal representations, called mode conversions, led to new
physical interpretations and ideas. However, mathematical expressions were
deduced by analogy with induced representation results leading thereby to some
inconsistencies.\cite{Smida 2000}

The purpose of the present work is to overcome the latter inconsistencies by
abandoning the fiber bundle structure in favor of an $\left(  M^{4}\times
M^{4}\right)  \otimes Z_{4}$, where $M^{4}$ is Minkowski space and $Z_{4}$ is
a discrete space with four elements. The inducing method is applied then
between symmetries of the same type only, and connection in the continuous
directions is taken into account. Transitions between symmetries of different
(external or internal) types is realized by means of connections in the
discrete direction.

The idea of discrete structure is drawn from works of Konisi\cite{Konisi 1996}
and Kubo\cite{Kubo 1998} who associated connections in the discrete directions
with Higgs fields without recourse to noncommutative geometry (NCG). For the
sake of definiteness, we mention that our work is neither to be compared with
that of Konisi and Kubo nor with those based on NCG. We have just used the
discrete structure to provide our idea of conversion with mathematical consistency.

In Sec. 2, the state spaces structure, the connections, and the physical
interpretations are presented. In Sec. 3, curvatures are calculated following
Kubo's work.\cite{Kubo 1998} In Sec. 4, propagators containing both types of
connections are deduced and Sec. 5 is devoted to the conclusion.

\section{The structure and connections}

To describe an extended particle composed of two local modes let us consider a
$\left(  M^{4}\times M^{4}\right)  \otimes Z_{4}$ structure, where $M^{4}$ is
Minkowski space-time and $Z_{4}$ is the discrete space with four elements.
States $\Psi^{z}$ of the particle belong to Hilbert spaces $H^{z}$ and are
considered as physical wave functions in the sense of providing all the
physical properties of the particle but not probabilities. The latter are
provided by functionals $X^{z}[\Psi^{z}]\left(  x,\xi\right)  =\psi\left(
x,\xi;z\right)  $.\cite{Destouches 1956} The variables $x$ and $\xi$ belong,
respectively, to the first and second space-time and the variable $z$ is an
element of $Z_{4}$ taking values $p$ for pure, $c$ for crossed, $e$ for
external, and $i$ for internal. Each case corresponds to a certain type of
localizability of the extended particle composed of a first mode localized at
$x$ and a second mode localized at $\xi$. In the pure case, the first mode is
localized in the external space-time and the second mode in the internal
space-time. The crossed case is the reverse of the former. The external and
internal cases mean that both modes are localized in external or internal
space-time, respectively. In other words, the localizability type $z$
attributes a fixed physical meaning to each space in the product $\left(
M^{4}\times M^{4}\right)  $ as being external or internal space-time. It
endows thereby the extended particle with the property of being completely
observable as a bilocal object in external space-time (external case),
partially observable as a local object in each space-time (pure and crossed
cases), or unobservable (internal case).

In the present work, we associate to each type of localizability $z$ a
symmetry group $G\left(  z\right)  $ with elements
\begin{equation}
U=\exp iT\left(  z\right)  .\theta\left(  x,\xi;z\right)
\end{equation}
We can assume that
\begin{equation}
T\left(  z\right)  .\theta\left(  x,\xi;z\right)  =T_{a}\left(  z\right)
\theta^{a}\left(  x,\xi;z\right)  +T_{\alpha}\left(  z\right)  \theta^{\alpha
}\left(  x,\xi;z\right)
\end{equation}
where $T_{a}$ and $\theta^{a}$ are, respectively, generators and parameters of
transformations related with the first space-time. In the same way,
$T_{\alpha}$ and $\theta^{\alpha}$ are generators and parameters of
transformations related with the second space-time.

These gauge transformations can be either spatiotemporal or unitary and induce
a connection corresponding to parallel transport in the continuous directions,
i.e. in one or both space-times. In fact, in the covariant
derivative\cite{Kubo 1998}
\begin{equation}
\nabla^{z}\psi\left(  x,\xi;z\right)  =\psi\left(  x+\delta x,\xi+\delta
\xi;z\right)  -\psi_{||}\left(  x+\delta x,\xi+\delta\xi;z\right)
\end{equation}
the parallel transported field $\psi_{||}\left(  x+\delta x,\xi+\delta
\xi;z\right)  $, from a location $\left(  x,\xi\right)  $ to another location
$\left(  x+\delta x,\xi+\delta\xi\right)  $, can be written
\begin{align}
\psi_{||}\left(  x+\delta x,\xi+\delta\xi;z\right)   & =H\left(  x+\delta
x,\xi+\delta\xi;z\right)  \psi\left(  x,\xi;z\right) \\
H\left(  x+\delta x,\xi+\delta\xi;z\right)   & =1-i\omega_{i}\left(
x,\xi;z\right)  \delta x^{i}-i\omega_{\mu}\left(  x,\xi;z\right)  \delta
\xi^{\mu}\label{HCI}%
\end{align}
The Lie algebra valued connection one-forms corresponding to gauge
transformations in each space-time are written in terms of gauge potentials
$A\left(  x,\xi;z\right)  $:
\begin{align}
\omega_{i}\left(  x,\xi;z\right)   & =T_{a}\left(  z\right)  A_{i}^{a}\left(
x,\xi;z\right) \\
\omega_{\mu}\left(  x,\xi;z\right)   & =T_{\alpha}\left(  z\right)  A_{\mu
}^{\alpha}\left(  x,\xi;z\right)
\end{align}
The covariant derivative takes then the following form
\begin{align}
\nabla^{z}  & =\delta x^{i}\nabla_{i}^{z}+\delta\xi^{\mu}\nabla_{\mu}^{z}\\
\nabla_{i}^{z}  & =\partial_{i}+i\omega_{i}\left(  x,\xi;z\right)
=\partial_{i}+iT_{a}\left(  z\right)  A_{i}^{a}\left(  x,\xi;z\right) \\
\nabla_{\mu}^{z}  & =\partial_{\mu}+i\omega_{\mu}\left(  x,\xi;z\right)
=\partial_{\mu}+iT_{\alpha}\left(  z\right)  A_{\mu}^{\alpha}\left(
x,\xi;z\right)
\end{align}
In the same manner, a covariant difference can be defined in the discrete
direction. Parallel transport is a transition from one type of localizability
(say $z^{\prime}$) to another ($z$)
\begin{equation}
\psi_{||}\left(  x,\xi;z\right)  =H\left(  x,\xi;z,z^{\prime}\right)
\psi\left(  x,\xi;z^{\prime}\right)
\end{equation}
Covariant difference is then written as follows
\begin{align}
\delta_{z^{\prime}}\psi\left(  x,\xi;z\right)   & =\psi\left(  x,\xi;z\right)
-\psi_{||}\left(  x,\xi;z\right) \\
& =\psi\left(  x,\xi;z\right)  -H\left(  x,\xi;z,z^{\prime}\right)
\psi\left(  x,\xi;z^{\prime}\right) \nonumber
\end{align}
but the $H\left(  x,\xi;z,z^{\prime}\right)  $ field has not a conventional
expression in terms of one forms and gauge potentials. It can be interpreted
as a transition operator from one type of localizability to another. It
corresponds then to a conversion of internal modes to external ones and
vice-versa. This conversion can be viewed as a creation of the particle when
it passes from an unobservable state to a partially or completely observable
one. It is viewed as an annihilation in the inverse transitions. It is then
natural to define the conjugate of such a conversion by the following
relation
\begin{equation}
H^{\dagger}\left(  x,\xi;z,z^{\prime}\right)  =H\left(  x,\xi;z^{\prime
},z\right)
\end{equation}

\section{Curvatures}

Now, we define and calculate different types of curvatures stemming from the
structure considered in this work.

The first type of curvature corresponds to parallel transport along closed
paths in the continuous direction and is given by the well known strength
field $F_{AB}\left(  x,\xi;z\right)  =-i[\nabla_{A}^{z},\nabla_{B}^{z}]$
components where the indices $A$ and $B$ take the values $i$ or $\mu$
\begin{equation}
F_{AB}\left(  x,\xi;z\right)  =\partial_{\lbrack A}\omega_{B]}\left(
x,\xi;z\right)  +i[\omega_{A}(x,\xi;z),\omega_{B}(x,\xi;z)]
\end{equation}
Subscript brackets $[,]$ indicates an antisymmetrization over the indices and
ordinary ones are commutator of connection forms. If parallel transport takes
place in the first space-time, the curvature takes the following form
\begin{equation}
F_{ij}\left(  x,\xi;z\right)  =\partial_{\lbrack i}\omega_{j]}\left(
x,\xi;z\right)  +i[\omega_{i}(x,\xi;z),\omega_{j}(x,\xi;z)]
\end{equation}
For a path in the second space-time, the curvature is
\begin{equation}
F_{\mu\nu}\left(  x,\xi;z\right)  =\partial_{\lbrack\mu}\omega_{\nu]}\left(
x,\xi;z\right)  +i[\omega_{\mu}(x,\xi;z),\omega_{\nu}(x,\xi;z)]
\end{equation}
and a closed path lying in the two spaces corresponds to the following
curvature
\begin{equation}
F_{i,\mu}\left(  x,\xi;z\right)  =\partial_{\lbrack i}\omega_{\mu]}\left(
x,\xi;z\right)  +i[\omega_{i}(x,\xi;z),\omega_{\mu}(x,\xi;z)]
\end{equation}
If the symmetry groups of each space-time commute, the latter curvature
becomes
\begin{equation}
F_{i,\mu}\left(  x,\xi;z\right)  =\partial_{\lbrack i}\omega_{\mu]}\left(
x,\xi;z\right)
\end{equation}
and, if in addition each connection depends only on its corresponding
space-time variable, this curvature vanishes identically.

The second type of curvature is concerned with a combination of a parallel
transport in the continuous direction with a parallel transport in the
discrete direction, Fig.(1).

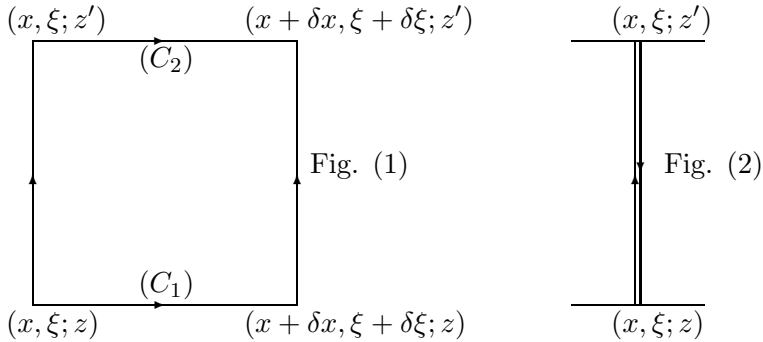
\begin{picture} (200,140)
\put(40,110) {\small{$(C_2)$}}
\put(40,25) {\small{$(C_1)$}}
\put(105,70) {\small{Fig. (1)}}
\put(-10,10) {\small{$(x,\xi;z)$}}
\put(-10,125) {\small{$(x,\xi;z')$}}
\put(80,125) {\small{$(x + \delta x,\xi+ \delta\xi;z')$}}
\put(80,10) {\small{$(x + \delta x,\xi+ \delta\xi;z)$}}
\put(0,20) {\vector(1,0){50}}
\put(0,20) {\vector(0,1){50}}
\put(0,120) {\vector(1,0){50}}
\put(100,20) {\vector(0,1){50}}
\put(50,20) {\line(1,0){50}}
\put(0,70) {\line(0,1){50}}
\put(50,120) {\line(1,0){50}}
\put(100,70) {\line(0,1){50}}
\end{picture}
\begin{picture} (150,140)
\put(15,10) {\small{$(x,\xi;z)$}}
\put(15,125) {\small{$(x,\xi;z')$}}
\put(35,70) {\small{Fig. (2)}}
\put(24,20) {\vector(0,1){50}}
\put(26,120) {\vector(0,-1){50}}
\put(0,20) {\line(1,0){50}}
\put(24,70) {\line(0,1){50}}
\put(0,120) {\line(1,0){50}}
\put(26,20) {\line(0,1){50}}
\end{picture}

The curvature is defined as a difference between two paths $C_{1}$ and $C_{2}
$ where
\begin{align}
C_{1}  & =H\left(  x+\delta x,\xi+\delta\xi;z^{\prime},z\right)  H\left(
x+\delta x,\xi+\delta\xi;z\right)  \psi\left(  x,\xi;z\right) \\
C_{2}  & =H\left(  x+\delta x,\xi+\delta\xi;z^{\prime}\right)  H\left(
x,\xi;z^{\prime},z\right)  \psi\left(  x,\xi;z\right)
\end{align}
We have
\begin{equation}
C_{1}-C_{2}=\{\delta x^{i}F_{iz^{\prime}}^{H}+\delta\xi^{\mu}F_{\mu z^{\prime
}}^{H}\}\psi\left(  x,\xi;z\right)
\end{equation}
where
\begin{align}
F_{iz^{\prime}}^{H}\left(  x,\xi;z\right)   & =\partial_{i}H\left(
x,\xi;z^{\prime},z\right) \\
& -iH\left(  x,\xi;z^{\prime},z\right)  \omega_{i}(x,\xi;z)+i\omega_{i}%
(x,\xi;z^{\prime})H\left(  x,\xi;z^{\prime},z\right) \nonumber\\
F_{\mu z^{\prime}}^{H}\left(  x,\xi;z\right)   & =\partial_{\mu}H\left(
x,\xi;z^{\prime},z\right) \\
& -iH\left(  x,\xi;z^{\prime},z\right)  \omega_{\mu}(x,\xi;z)+i\omega_{\mu
}(x,\xi;z^{\prime})H\left(  x,\xi;z^{\prime},z\right) \nonumber
\end{align}
It is clear that if parallel transport in the continuous direction concerns
one space-time, only the corresponding curvature has to be considered.

Parallel transport of the third type curvature links two points in the
discrete direction only, Fig. (2). Then
\begin{equation}
F_{z^{\prime}(z)}\left(  x,\xi;z\right)  =1-H\left(  x,\xi;z,z^{\prime
}\right)  H\left(  x,\xi;z^{\prime},z\right) \label{f2z}%
\end{equation}
For purely discrete curvatures, we adopt the following notation. The initial
point $z$ in the diagram is considered as an argument, the intermediate point
$z^{\prime}$ as an index, and the end point $z$ as an index between parenthesis.

There are also parallel transports linking three and four points in the
discrete directions depicted in Figs. (3) and (4), respectively.

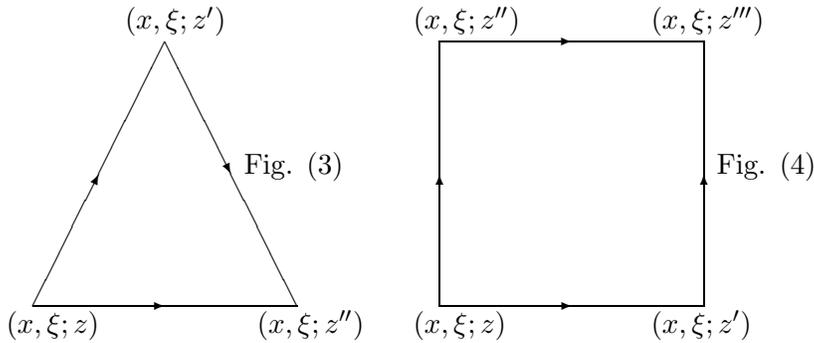
\begin{picture} (150,140)
\put(-10,10) {\small{$(x,\xi;z)$}}
\put(35,125) {\small{$(x,\xi;z')$}}
\put(85,10) {\small{$(x,\xi;z'')$}}
\put(80,70) {\small{Fig. (3)}}
\put(25,70) {\line(1,2) {25}}
\put(75,70) {\line(1,-2) {25}}
\put(100,20) {\line(-1,0) {50}}
\put(0,20) {\vector(1,2) {25}}
\put(50,120) {\vector(1,-2) {25}}
\put(0,20) {\vector(1,0) {50}}
\end{picture}
\begin{picture} (150,140)
\put(-10,10) {\small{$(x,\xi;z)$}}
\put(-10,125) {\small{$(x,\xi;z'')$}}
\put(80,125) {\small{$(x,\xi;z''')$}}
\put(80,10) {\small{$(x,\xi;z')$}}
\put(105,70) {\small{Fig. (4)}}
\put(0,20) {\vector(1,0){50}}
\put(0,20) {\vector(0,1){50}}
\put(0,120) {\vector(1,0){50}}
\put(100,20) {\vector(0,1){50}}
\put(50,20) {\line(1,0){50}}
\put(0,70) {\line(0,1){50}}
\put(50,120) {\line(1,0){50}}
\put(100,70) {\line(0,1){50}}
\end{picture}

They give the curvature of the third type which has the following form
\begin{equation}
F_{z^{\prime}(z^{\prime\prime})}\left(  x,\xi;z\right)  =H\left(
x,\xi;z^{\prime\prime},z\right)  -H\left(  x,\xi;z^{\prime\prime},z^{\prime
}\right)  H\left(  x,\xi;z^{\prime},z\right) \label{f3z}%
\end{equation}
and the curvature of the fourth type which has an analogous expression
\begin{equation}
F_{z^{\prime}z^{\prime\prime}(z^{\prime\prime\prime})}\left(  x,\xi;z\right)
=H\left(  x,\xi;z^{\prime\prime\prime},z^{\prime\prime}\right)  H\left(
x,\xi;z^{\prime\prime},z\right)  -H\left(  x,\xi;z^{\prime\prime\prime
},z^{\prime}\right)  H\left(  x,\xi;z^{\prime},z\right) \label{f4z}%
\end{equation}
Note that curvature (\ref{f2z}) is compatible with (\ref{f3z}) since $H\left(
x,\xi;z,z\right)  =1$ and that (\ref{f4z}) can be derived from (\ref{f3z})
\begin{equation}
F_{z^{\prime}z^{\prime\prime}(z^{\prime\prime\prime})}\left(  x,\xi;z\right)
=F_{z^{\prime\prime}(z^{\prime\prime\prime})}(z)-F_{z^{\prime}(z^{\prime
\prime\prime})}(z)
\end{equation}
The latter relation shows antisymmetry with respect to $z^{\prime}$ and
$z^{\prime\prime}$. Moreover, it is easy to show that
\begin{equation}
F_{\bullet(z^{\prime})}\left(  x,\xi;z\right)  =F_{\bullet(z)}^{\dagger
}\left(  x,\xi;z^{\prime}\right)
\end{equation}
where the dot ($\bullet$) is to be replaced by the adequate arguments of the
purely discrete curvatures. Consequently, curvature (\ref{f2z}) is hermitic.

All our curvatures are analogous to those of local theories\cite{Konisi
1996,Kubo 1998} except the continuous curvature which contains extra terms due
to extension.

\section{Propagators}

\subsection{Free case}

The free case has already been studied\cite{Smida 2000} by considering a field
$\psi\left(  x,\xi\right)  $ with an external space-time variable $x$, an
internal space-time variable $\xi$, and without the $z$ variable (i.e. on
$M^{4}\times M^{4}$ with $z=p$). The latter field is quantized by applying the
method of induced representations to the symmetries of both space-times
independently. For that purpose, the momentum representation with functionals
$\varphi\left(  v,\zeta\right)  $ was considered. According to the induced
representation scheme, transitions from a localized state (configuration
representation) to another take place via momentum states; the transition from
configuration space to momentum space has been called a materialization
$\mathcal{K}$, the inverse transition is a localization $\mathfrak{I}$.
Consideration of all materializations and all localizations between external
and internal spaces can be represented by the following diagram for the two
modes:
\begin{align*}
&  \underset{\left(  \alpha\right)  }{\text{Mode 1}:configuration}%
\overset{\mathcal{K}_{m_{\beta}}^{\beta\alpha}}{\rightarrow}\underset{\left(
\beta\right)  }{momentum}\overset{\mathfrak{I}_{m_{\beta}}^{\gamma\beta}%
}{\rightarrow}\underset{\left(  \gamma\right)  }{configuration}\\
&  \underset{\left(  \alpha^{\prime}\right)  }{\text{Mode 2}:configuration}%
\overset{\mathcal{K}_{\mu_{\beta^{\prime}}}^{\beta^{\prime}\alpha^{\prime}}%
}{\rightarrow}\underset{\left(  \beta^{\prime}\right)  }{momentum}%
\overset{\mathfrak{I}_{\mu_{\beta^{\prime}}}^{\gamma^{\prime}\beta^{\prime}}%
}{\rightarrow}\underset{\left(  \gamma^{\prime}\right)  }{configuration}%
\end{align*}
In the beginning \cite{Smida 1995}, we considered the case the $z=p$ where the
first line of the diagram contained external spaces only ($\alpha=\beta
=\gamma=external$), and the second line contained internal spaces only
($\alpha^{\prime}=\beta^{\prime}=\gamma^{\prime}=internal$). The masses
$m_{\beta}$ and $\mu_{\beta^{\prime}}$ take values $\mu$ or $m$ depending on
whether the momentum space is internal or external, respectively.

Afterwards, we have assumed that the first external mode may materialize or
localize in the internal space, and vice versa for the second mode \cite{Smida
2000}. Consequently, the indices $\alpha,\alpha^{\prime},\beta,\beta^{\prime
},\gamma,\gamma^{\prime}$ can take any of the values \textit{external }or
\textit{internal,} independently, so that we have 64 possibilities. A
propagation corresponding to a fixed combination of the values of the six
indices has been deduced by the method of induced representations \cite{Smida
2000}
\begin{equation}
\psi\left(  x_{\gamma},\xi_{\gamma^{\prime}}\right)  =\int dx_{\alpha}\int
d\xi_{\alpha^{\prime}}\Pi_{m_{\beta}}^{\mu_{\beta^{\prime}}}\left(  x_{\gamma
},\xi_{\gamma^{\prime}};x_{\alpha},\xi_{\alpha^{\prime}}\right)  \psi\left(
x_{\alpha},\xi_{\alpha^{\prime}}\right)
\end{equation}
and has been written as product of two point-like propagations
\begin{equation}
\Pi_{m_{\beta}}^{\mu_{\beta^{\prime}}}\left(  x_{\gamma},\xi_{\gamma^{\prime}%
};x_{\alpha},\xi_{\alpha^{\prime}}\right)  =\Pi_{m_{\beta}}\left(  x_{\gamma
},x_{\alpha}\right)  \Pi^{\mu_{\beta^{\prime}}}\left(  \xi_{\gamma^{\prime}%
},\xi_{\alpha^{\prime}}\right)
\end{equation}
The point-like propagator has the following form \cite{Mensky 1976}
\begin{equation}
\Pi_{m_{\beta}}^{\pm}\left(  x_{\gamma},x_{\alpha}\right)  =\frac{m_{\beta
}^{2}}{2\left(  2\pi\right)  ^{3}}\int_{C^{1}}dv\exp\mp i\left[  m_{\beta
}v_{\beta}(x_{\gamma}-x_{\alpha})\right]
\end{equation}
where the velocity $v_{\beta}$ belongs to the positive sheet $C^{1}$ of the
unit mass hyperboloid with invariant measure $dv$. The signs $\left(
+\right)  $ and $\left(  -\right)  $ are explicitly shown here and correspond
to modes and antimodes, respectively. An identical expression holds for
$\Pi^{\mu_{\beta^{\prime}}}\left(  \xi_{\gamma^{\prime}},\xi_{\alpha^{\prime}%
}\right)  $ with mass $\mu_{\beta^{\prime}}$ and velocity $\zeta
_{\beta^{\prime}}$. The product form of the extended particle propagator was
derived by an analogy with induced representation. This is not consistent
since it describes transitions between external and internal spaces of one
mode with no interaction. Moreover, these same crossed transitions makes the
first fiber bundle structure inadequate \cite{Smida 2000}. Using the $\left(
M^{4}\times M^{4}\right)  \otimes Z_{4}$ structure, we can get rid of these
problems. We begin by considering the discrete connection first.

\subsection{Discrete connection case}

In the first of our previous works,\cite{Smida 1995} interaction has been
considered in the inducing scheme with no crossed transitions. We have shown
that the localization, materialization and propagation processes are locally
affected by parallel transport operators containing the interaction. The
latter operators are path ordered exponentials, the analogues of the $H$
mappings in continuous directions. And since we have already remarked that the
operator $H\left(  x,\xi;z,z^{\prime}\right)  $ realizing a discrete
transition from $z^{\prime}$ to $z$ is a parallel transport operator, let us
use that recipe in the context of the present work. Materialization of a type
$z^{\prime}$ localized state into a type $z$ real state can then be written
as
\begin{equation}
\varphi\left(  v,\zeta,z\right)  =\left[  \mathcal{K}^{z}H\left(  z,z^{\prime
}\right)  \psi\right]  \left(  v,\zeta,z\right) \label{Mat}%
\end{equation}
Materialization operators $\mathcal{K}^{z}$ are products of materializations
of the first and second modes. Consider, for instance, the case where we have
a pure localized initial state $z^{\prime}=p$ and a crossed material final
state $z=c$, the materialization is then
\begin{equation}
\mathcal{K}^{c}H\left(  c,p\right)  =\mathcal{K}_{\mu0}\mathcal{K}%
_{m0}H\left(  c,p\right)
\end{equation}
The rule is the following: mass and spin ($0$ here) parameters are determined
by the real state. In the crossed state $\varphi\left(  v,\zeta,c\right)  $,
the first mode is internal and the second external. Hence, the first operator
$\mathcal{K}_{\mu0}$ carries the internal mass $\mu$ and intertwines
representations of the first group. The second operator $\mathcal{K}_{m0}$
carries the external mass $m$ and concerns the second group. In integral form,
we have
\begin{align}
\varphi\left(  v,\zeta,c\right)   &  =\frac{m\mu}{2(2\pi)^{3}}\int_{M^{4}%
}dx\int_{M^{4}}d\xi\exp\pm i(\mu vx+m\zeta\xi)\times\nonumber\\
&  H\left(  x,\xi;c,p\right)  \psi\left(  x,\xi,p\right)
\end{align}
This relation can be written in the general case
\begin{align}
\varphi\left(  v,\zeta,z\right)   &  =\frac{m_{z}\mu_{z}}{2(2\pi)^{3}}%
\int_{M^{4}}dx\int_{M^{4}}d\xi\exp\pm i(m_{z}vx+\mu_{z}\zeta\xi)\times
\nonumber\\
&  H\left(  x,\xi;z,z^{\prime}\right)  \psi\left(  x,\xi,z^{\prime}\right)
\label{MatH}%
\end{align}
where $m_{z}$ and $\mu_{z}$ are masses of the first and second modes,
respectively. The nature $z$ of the real state specifies these masses (see
Table \ref{table1} after the conclusion).

The mappings $H$ have been defined in the configuration representation, but it
is clear that analogous mappings $\widetilde{H}$ exist in the momentum
representation on $\left(  C^{1}\otimes C^{1}\right)  \otimes Z_{4}$.
Consequently, we can write relation (\ref{Mat}) in the form
\begin{equation}
\varphi\left(  v,\zeta,z\right)  =\left[  \widetilde{H}\left(  z,z^{\prime
}\right)  \mathcal{K}^{z^{\prime}}\psi\right]  \left(  v,\zeta,z\right)
\end{equation}
In this case the integral form becomes
\begin{align}
\varphi\left(  v,\zeta,z\right)   &  =\frac{m_{z^{\prime}}\mu_{z^{\prime}}%
}{2(2\pi)^{3}}\widetilde{H}\left(  v,\zeta;z,z^{\prime}\right)  \times
\nonumber\\
&  \int_{M^{4}}dx\int_{M^{4}}d\xi\exp\pm i(m_{z^{\prime}}vx+\mu_{z^{\prime}%
}\zeta\xi)\psi\left(  x,\xi,z^{\prime}\right)
\end{align}
Compatibility between the two expressions imposes the following relation
between materializations on the sheet $z$ :
\begin{equation}
\widetilde{H}\left(  z,z^{\prime}\right)  \mathcal{K}^{z^{\prime}}%
=\mathcal{K}^{z}H\left(  z,z^{\prime}\right)
\end{equation}

In the same way, localization of a real state $z$ into a localized state
$z^{\prime}$, is written as
\begin{equation}
\psi\left(  x,\xi,z^{\prime}\right)  =\left[  H\left(  z^{\prime},z\right)
\mathfrak{I}^{z}\varphi\right]  \left(  x,\xi,z^{\prime}\right)
\end{equation}
The integral form is
\begin{align}
\psi\left(  x,\xi,z^{\prime}\right)   &  =H\left(  x,\xi;z^{\prime},z\right)
\frac{m_{z}\mu_{z}}{2(2\pi)^{3}}\int_{C^{1}}dv\int_{C^{1}}d\zeta
\times\nonumber\\
&  \exp\mp i(m_{z}vx+\mu_{z}\zeta\xi)\varphi\left(  v,\zeta,z\right)
\label{LocH}%
\end{align}
If we begin by a transition from sheet $z$ to sheet $z^{\prime}$ in the
momentum representation before realizing the localization, we get
\begin{equation}
\psi\left(  x,\xi,z^{\prime}\right)  =\left[  \mathfrak{I}^{z^{\prime}%
}\widetilde{H}\left(  z^{\prime},z\right)  \varphi\right]  \left(
x,\xi,z^{\prime}\right)
\end{equation}
and
\begin{align}
\psi\left(  x,\xi,z^{\prime}\right)   &  =\frac{m_{z^{\prime}}\mu_{z^{\prime}%
}}{2(2\pi)^{3}}\int_{C^{1}}dv\int_{C^{1}}d\zeta\exp\mp i(m_{z^{\prime}}%
vx+\mu_{z^{\prime}}\zeta\xi)\times\nonumber\\
&  \widetilde{H}\left(  v,\zeta;z^{\prime},z\right)  \varphi\left(
v,\zeta,z\right)
\end{align}
The compatibility relation being then
\begin{equation}
\mathfrak{I}^{z^{\prime}}\widetilde{H}\left(  z^{\prime},z\right)  =H\left(
z^{\prime},z\right)  \mathfrak{I}^{z}%
\end{equation}

According to the inducing method, the propagation operator is obtained by
composing a materialization and a localization. The new processes of
materialization and localization (containing transitions from a sheet of
$M^{4}\otimes M^{4}\otimes Z_{4}$ to another) are combined in the same way.
Because of the compatibility relations, the propagator can be deuced in
several equivalent ways:
\begin{subequations}
\begin{align}
\Pi^{z\pm}\left(  z^{\prime\prime},z^{\prime}\right)   &  =H\left(
z^{\prime\prime},z\right)  \mathfrak{I}^{z}\mathcal{K}^{z}H\left(
z,z^{\prime}\right) \\
&  =\mathfrak{I}^{z^{\prime\prime}}\widetilde{H}\left(  z^{\prime\prime
},z\right)  \widetilde{H}\left(  z,z^{\prime}\right)  \mathcal{K}^{z^{\prime}%
}\\
&  =H\left(  z^{\prime\prime},z\right)  \mathfrak{I}^{z}\widetilde{H}\left(
z,z^{\prime}\right)  \mathcal{K}^{z^{\prime}}\\
&  =\mathfrak{I}^{z^{\prime\prime}}\widetilde{H}\left(  z^{\prime\prime
},z\right)  \mathcal{K}^{z}H\left(  z,z^{\prime}\right)
\end{align}
and corresponds to a transition from a type $z^{\prime}$ localized state to a
type $z^{\prime\prime}$ localized state via a type $z$ real state. The kernel
of the propagation
\end{subequations}
\begin{equation}
\psi\left(  x^{\prime\prime},\xi^{\prime\prime},z^{\prime\prime}\right)
=\int_{M^{4}}dx^{\prime}\int_{M^{4}}d\xi^{\prime}\Pi^{z\pm}\left(
x^{\prime\prime},\xi^{\prime\prime},z^{\prime\prime};x^{\prime},\xi^{\prime
},z^{\prime}\right)  \psi\left(  x^{\prime},\xi^{\prime},z^{\prime}\right)
\end{equation}
corresponding to the first expression of the propagator $\Pi^{z}\left(
z^{\prime\prime},z^{\prime}\right)  $ has then the following form
\begin{align}
\Pi^{z\pm}\left(  x^{\prime\prime},\xi^{\prime\prime},z^{\prime\prime
};x^{\prime},\xi^{\prime},z^{\prime}\right)   &  =H\left(  x^{\prime\prime
},\xi^{\prime\prime};z^{\prime\prime},z\right)  \left[  \frac{m_{z}\mu_{z}%
}{2(2\pi)^{3}}\right]  ^{2}\int_{C^{1}}dv\int_{C^{1}}d\zeta\times\\
&  \exp\mp i\left[  m_{z}v(x^{\prime\prime}-x^{\prime})+\mu_{z}\zeta
(\xi^{\prime\prime}-\xi^{\prime})\right]  H\left(  x^{\prime},\xi^{\prime
};z,z^{\prime}\right) \nonumber
\end{align}
The three other expressions are analogous with adequate mappings $H$.

For particles with non vanishing spin, the propagator is
\begin{align}
\Pi^{z\pm}\left(  x^{\prime\prime},\xi^{\prime\prime},z^{\prime\prime
};x^{\prime},\xi^{\prime},z^{\prime}\right)   &  =H\left(  x^{\prime\prime
},\xi^{\prime\prime};z^{\prime\prime},z\right)  \left[  \frac{m_{z}\mu_{z}%
}{2(2\pi)^{3}}\right]  ^{2}\int_{C^{1}}dv\int_{C^{1}}d\zeta S^{z\pm}\left(
v,\zeta\right)  \times\\
&  \exp\mp i\left[  m_{z}v(x^{\prime\prime}-x^{\prime})+\mu_{z}\zeta
(\xi^{\prime\prime}-\xi^{\prime})\right]  H\left(  x^{\prime},\xi^{\prime
};z,z^{\prime}\right) \nonumber
\end{align}
where \cite{Mensky 1976}
\begin{align}
S^{z\pm}\left(  v,\zeta\right)   &  =S_{j_{z}}^{\pm}\left(  v\right)
S_{\sigma_{z}}^{\pm}\left(  \zeta\right) \\
S_{j_{z}}^{\pm}\left(  v\right)   &  =D\left(  v_{L}\right)  \mathfrak{I}%
_{j_{z}}^{\pm}\mathcal{K}_{j_{z}}^{\pm}D\left(  v_{L}^{-1}\right) \\
S_{\sigma_{z}}^{\pm}\left(  \zeta\right)   &  =D\left(  \zeta_{L}\right)
\mathfrak{I}_{\sigma_{z}}^{\pm}\mathcal{K}_{\sigma_{z}}^{\pm}D\left(
\zeta_{L}^{-1}\right)
\end{align}
In the $D\left(  v_{L}\right)  \mathfrak{I}_{j_{z}}^{\pm}$ and $\mathcal{K}%
_{j_{z}}^{\pm}D\left(  v_{L}^{-1}\right)  $operators, corresponding to the
first momentum space in $\left(  C^{1}\otimes C^{1}\right)  \otimes Z_{4}$,
the $D$ matrix is a representation of the Lorentz subgroup, $v_{L}$ is the
velocity boost, and the operators $\mathfrak{I}_{j_{z}}^{\pm}$ and
$\mathcal{K}_{j_{z}}^{\pm}$ are constant matrices verifying $\mathcal{K}%
_{j_{z}}^{\pm}\mathfrak{I}_{j_{z}}^{\pm}=1$ and whose dimension depends on the
spin $j_{z}$ ($\mathfrak{I}_{0}^{\pm}$ $=$ $\mathcal{K}_{0}^{\pm}=1$).
Quantities $D\left(  \zeta_{L}\right)  \mathfrak{I}_{\sigma_{z}}^{\pm}$ and
$\mathcal{K}_{\sigma_{z}}^{\pm}D\left(  \zeta_{L}^{-1}\right)  $corresponding
to the second momentum space commute with those of the first space. Let us
denote the external mass and spin by $\left(  m,j\right)  $ and the internal
ones by $\left(  \mu,\sigma\right)  $, we then get masses and spins in the
expressions of materialization, localization, and propagation with a momentum
state $\varphi\left(  v,\zeta,z\right)  $ as shown in Table \ref{table1} after
the conclusion.

\subsection{Discrete and continuous connection case}

We first consider the case of one sheet with fixed $z$. The propagator has
already been deduced for this case (which is equivalent to $\left(
M^{4}\times M^{4}\right)  $) by means of trajectory semigroups induced
representations in each space-time \cite{Smida 1998}. Use of trajectory
semigroups is imposed as far as the continuous connection is to be taken into
account in the inducing method of quantization.

A trajectory in $M^{4}$ is a class of parallel curves $x(l)$. It is
represented by an element
\begin{equation}
\lbrack u]_{t}=\{u(l)\in\mathbf{R}^{4}/\,0\leq l\leq
t\},\;\;\;\;\;\;t=(s-s^{\prime}),\;\;\;\;\;\;u(l)=(\frac{dx}{dl})_{s^{\prime
}+l}%
\end{equation}
where $x^{\prime}=x(s^{\prime})$ and $x=x(s)$ are the initial and final points
of the curve $x\left(  l\right)  $, respectively. Right action of trajectories
on space-time points is defined in the following way
\begin{equation}
x[u]_{t}=x^{\prime}=x-\int_{0}^{t}u(l)\,dl\label{ex}%
\end{equation}
Trajectories $[\gamma]_{\tau}$ are also defined for curves $\xi\left(
\lambda\right)  $ of the second Minkowski space-time in $\left(  M^{4}\times
M^{4}\right)  \otimes Z_{4}$. To each pair of trajectories is associated a
translation operator $U([u]_{t},[\gamma]_{\tau})$
\begin{equation}
\lbrack U([u]_{t},[\gamma]_{\tau})\psi](x,\xi;z)=\psi(x[u]_{t},\xi
\lbrack\gamma]_{\tau};z)
\end{equation}
and a parallel transport operator $H_{([u]_{t},[\gamma]_{\tau};z)}$, acting
jointly with $U$ and taking account of the continuous gauge fields effect
\begin{equation}
\lbrack H_{([u]_{t},[\gamma]_{\tau};z)}U([u]_{t},[\gamma]_{\tau})\psi
](x,\xi;z)=H_{([u]_{t},[\gamma]_{\tau})}(x,\xi;z)\psi(x[u]_{t},\xi
\lbrack\gamma]_{\tau};z)
\end{equation}
The connection $H_{([u]_{t},[\gamma]_{\tau})}(x,\xi;z)$ corresponds to a
parallel transport from $(x[u]_{t},\xi\lbrack\gamma]_{\tau};z)$ to $(x,\xi;z)$
along two curves belonging to trajectories $[u]_{t}$ and $[\gamma]_{\tau}$.
Its infinitesimal form is given by relation (\ref{HCI}) and its finite form is
an ordered path integral over both curves
\begin{equation}
H(x[u]_{t},\xi\lbrack\gamma]_{\tau};z)=P\left[  \exp-\int_{(x[u]_{t}%
,\xi\lbrack\gamma]_{\tau};z)}^{(x,\xi;z)}\left(  i\omega_{i}\left(
x,\xi;z\right)  dx^{i}+i\omega_{\mu}\left(  x,\xi;z\right)  d\xi^{\mu}\right)
\right]
\end{equation}
In our previous works, $i\omega_{i}\left(  x,\xi;z\right)  dx^{i}$ has been
denoted $\Gamma_{i}\left(  x\right)  dx^{i}$ whereas $i\omega_{\mu}\left(
x,\xi;z\right)  d\xi^{\mu}$ was not considered since we used a gauging of the
internal symmetry with respect to the external space-time only. The situation
is quite different now, we have a symmetry of two space-times gauged with
respect to both.

The one sheet propagation operator of functions $\psi\left(  x,\xi,z\right)  $
is a path integral expression\cite{Smida 1998} ($\theta$ is the step
function)
\begin{align}
\Pi_{z}^{c}  & =\int dt\,\theta(t)\,\exp(-im_{z}^{2}t)\int d\tau\,\theta
(\tau)\,\exp(-i\mu_{z}^{2}\tau)\int d[u]_{t}\int d[\gamma]_{\tau}\\
& \ \ \ \exp(\frac{-i}{4}\int_{0}^{t}dl\,u^{2}(l))\,\,\exp(\frac{-i}{4}%
\int_{0}^{\tau}d\lambda\,\gamma^{2}(\lambda))H_{([u]_{t},[\gamma]_{\tau}%
;z)}\,U([u]_{t},[\gamma]_{\tau})\nonumber
\end{align}
with measures
\begin{equation}
d[u]_{t}=\prod_{l=0}^{t}du(l)\,\,\,\,\,\,\,\,\,\,\,\,\,\,\,\,\,\,\,d[\gamma
]_{\tau}=\prod_{\lambda=0}^{\tau}d\gamma(\lambda)
\end{equation}
Now, we come to the implementation of our idea of conversion and proceed by
comparison with the case where the continuous connection is ignored, that is
when groups are used instead of semigroups. A general propagation amounts to a
transition from $\psi\left(  x^{\prime},\xi^{\prime};z^{\prime}\right)  $ to
$\psi\left(  x^{\prime\prime},\xi^{\prime\prime};z^{\prime\prime}\right)  $
through momentum representation in a $z$-sheet. The function $\psi\left(
x^{\prime},\xi^{\prime};z^{\prime}\right)  $ is transformed by means of
$H\left(  x^{\prime},\xi^{\prime};z,z^{\prime}\right)  $ and propagation is
realized from $\left(  x^{\prime},\xi^{\prime};z\right)  $ to $\left(
x^{\prime\prime},\xi^{\prime\prime};z\right)  $ in the $z$-sheet. Then the
result is transformed by $H\left(  x,\xi;z^{\prime\prime},z\right)  $. When
the continuous connection is considered (trajectory case), the two $H$ fields
representing the discrete connection must be included in the propagator and
masses have to be labeled with the intertwining sheet parameter $z$.

Hence, the most general expression of the propagator operator is
\begin{align}
\Pi_{z}^{c}\left(  z^{\prime\prime},z^{\prime}\right)   & =\int dt\,\theta
(t)\,\exp(-im_{z}^{2}t)\int d\tau\,\theta(\tau)\,\exp(-i\mu_{z}^{2}\tau)\\
& \int d[u]_{t}\int d[\gamma]_{\tau}\exp(\frac{-i}{4}\int_{0}^{t}%
dl\,u^{2}(l))\,\,\exp(\frac{-i}{4}\int_{0}^{\tau}d\lambda\,\gamma^{2}%
(\lambda))\nonumber\\
& H_{([u]_{t},[\gamma]_{\tau};z)}\left(  z^{\prime\prime},z^{\prime}\right)
\,U([u]_{t},[\gamma]_{\tau})\nonumber
\end{align}
The new operators $H_{([u]_{t},[\gamma]_{\tau};z)}\left(  z^{\prime\prime
},z^{\prime}\right)  $ defined by
\begin{align}
& \left[  H_{([u]_{t},[\gamma]_{\tau};z)}\left(  z^{\prime\prime},z^{\prime
}\right)  U([u]_{t},[\gamma]_{\tau}\psi\right]  (x,\xi;z^{\prime
})\,\nonumber\\
& =H\left(  x,\xi;z^{\prime\prime},z\right)  H_{([u]_{t},[\gamma]_{\tau}%
)}(x,\xi;z)\times\\
& H(x[u]_{t},\xi\lbrack\gamma]_{\tau};z,z^{\prime})\psi(x[u]_{t},\xi
\lbrack\gamma]_{\tau};z^{\prime})\nonumber
\end{align}
contain continuous and discrete connections. We note that the general
propagation operator is compatible with the one sheet propagation operator
since $H(x,\xi;z,z)=1$. In total, we have sixty-four propagations differing by
the values of $\left(  z^{\prime\prime},z,z^{\prime}\right)  $. Propagation of
fields can be written in the following way
\begin{align}
\psi(x^{\prime\prime},\xi^{\prime\prime};z^{\prime\prime})  & =[\Pi_{z}%
^{c}\left(  z^{\prime\prime},z^{\prime}\right)  \psi](x^{\prime\prime}%
,\xi^{\prime\prime};z^{\prime})\label{PGa}\\
& =\int dx^{\prime}d\xi^{\prime}\Pi_{z}^{c}\left(  x^{\prime\prime}%
,\xi^{\prime\prime},z^{\prime\prime};x^{\prime},\xi^{\prime},z^{\prime
}\right)  \psi(x^{\prime},\xi^{\prime};z^{\prime})\label{PGb}%
\end{align}
The kernel $\Pi_{z}^{c}\left(  x^{\prime\prime},\xi^{\prime\prime}%
,z^{\prime\prime};x^{\prime},\xi^{\prime},z^{\prime}\right)  $ in relation
(\ref{PGb}) is to be determined after calculation of (\ref{PGa}). Such a
kernel is interpreted as a spatiotemporal evolution of two modes from $\left(
x^{\prime},\xi^{\prime}\right)  $ to $\left(  x^{\prime\prime},\xi
^{\prime\prime}\right)  $, which may be accompanied by conversions (if $z$ is
different from $z^{\prime}$ or $z^{\prime\prime}$).

\section{Conclusion}

The $\left(  M^{4}\times M^{4}\right)  \otimes Z_{n}$ structure seemed
interesting in interpreting the geometrical origin of Higgs fields without
recourse to noncommutative geometry.\cite{Konisi 1996,Kubo 1998}

The present work reveals another aspect of this structure, which is not
concerned with the Higgs phenomenon. It opens the way to the construction of a
theory of extended particles interacting with gauge fields and reaches the
determination of a path integral form of the propagators. This work explains,
in a manner analogous to the Dirac sea for fermions, the creation of particles
by admitting their prior existence in unobservable states and the possibility
of their transition to observable ones. The $\left(  M^{4}\times M^{4}\right)
\otimes Z_{4}$ interest is that it provides the mathematical objects
representing these transitions, namely the discrete connections $H\left(
x,\xi;z,z^{\prime}\right)  $. Moreover, the theory allows consideration of
gauge fields corresponding to a symmetry localized not only with respect to
one space-time (generally, external space-time) but in both space-times.

Symmetries, connections, curvatures, and propagators have been presented in
their most general form. Propagators incorporate conversions and effects of
gauge fields in the continuous directions. Hence, the particle evolves in
space-time and this evolution may be accompanied by its annihilation, its
creation, or a transformation of those of its properties which are coupled to
the continuous gauge field.

The following step is the adoption of specific physical models and the
derivation of equation of motion for $\psi\left(  x,\xi;z\right)  $, $A\left(
x,\xi;z\right)  $, and $H\left(  x,\xi;z,z^{\prime}\right)  $. The study of
this question has already been initiated by the determination of curvatures
which may lead to a Lagrangian formulation. However, this is to be carefully
analyzed since the validity of a bilocal Lagrangian theory is not well
established. 
\begin{table}[h]
\begin{tabular}
[c]{|c|c|c|c|c|}\hline
$z=$ & $p$ (pure) & $c\ $(crossed) & $e\ $(external) & $i$ (internal)\\\hline
mode 1: $\left(  m_{z},j_{z}\right)  =$ & $\left(  m,j\right)  $ & $\left(
\mu,\sigma\right)  $ & $\left(  m,j\right)  $ & $\left(  \mu^{\prime}%
,\sigma^{\prime}\right)  $\\\hline
mode 2: $\left(  \mu_{z},\sigma_{z}\right)  =$ & $\left(  \mu,\sigma\right)  $
& $\left(  m,j\right)  $ & $\left(  m^{\prime},j^{\prime}\right)  $ & $\left(
\mu,\sigma\right)  $\\\hline
\end{tabular}
\caption{Notation of mass and spin according to momentum state z}%
\label{table1}
\end{table}

\textbf{Acknowledgment :} A. S., M. H., and A.-H. H. are supported by research
Project N${{}^{\circ}}$ : D 1602/06/2000.

\end{document}